\documentclass[12pt]{iopart}
\usepackage{cases}
\usepackage{dsfont}
\usepackage{xspace}
\usepackage{comment}

\newcommand{\latin}[1]{{\it #1}}
\newcommand{\ie}{\latin{i.e.}\@\xspace}
\newcommand{\gpvec}[1]{\mathbf{#1}}
\newcommand{\xvec}{\gpvec{x}}
\newcommand{\elabel}[1]{\label{#1}}
\newcommand{\ave}[1]{\left\langle #1 \right\rangle}
\newcommand{\spave}[1]{\overline{#1}}
\newcommand{\dint}[1]{\mathchoice{\!\mathrm{d}#1\,}{\!\mathrm{d}#1\,}{\!\mathrm{d}#1\,}{\!\mathrm{d}#1\,}}
\newcommand{\ddint}[1]{\mathchoice{\!\mathrm{d}^d#1\,}{\!\mathrm{d}^d#1\,}{\!\mathrm{d}^d#1\,}{\!\mathrm{d}^d#1\,}}
\newcommand{\GC}{\mathcal{G}}
\newcommand{\OC}{\mathcal{O}}
\newcommand{\GCtilde}{\tilde{\GC}}
\newcommand{\Gtilde}{\tilde{G}}
\newcommand{\utilde}{\tilde{u}}
\newcommand{\wtilde}{\tilde{w}}
\newcommand{\ytilde}{\tilde{y}}
\newcommand{\phitilde}{\tilde{\phi}}
\newcommand{\braket}[2]{\left\langle#1|#2\right\rangle}
\newcommand{\half}{\mathchoice{\frac{1}{2}}{(1/2)}{\frac{1}{2}}{(1/2)}}
\newcommand{\fourth}{\mathchoice{\frac{1}{4}}{(1/4)}{\frac{1}{4}}{(1/4)}}

\newcommand{\quarter}{\fourth}
\newcommand{\imag}{\imath}
\newcommand{\gpset}[1]{\mathds{#1}}

\newcommand{\canetset}[1]{{\mathchoice {\hbox{$\sf\textstyle #1\kern-0.4em #1$}}
{\hbox{$\sf\textstyle #1\kern-0.4em #1$}}
{\hbox{$\sf\scriptstyle #1\kern-0.3em #1$}}
{\hbox{$\sf\scriptscriptstyle #1\kern-0.2em #1$}}}}

\newcommand{\Nset}{\gpset{N}}

\newcommand{\Rset}{\gpset{R}}

\newcommand{\physical}{\text{\tiny physical}}

\newcommand{\dynamicalExpo}{z}
\newcommand{\roughnessExpo}{\alpha}
\newcommand{\rougheningExpo}{\beta}
\newcommand{\text}[1]{\textrm{\ #1\ }}
\newcommand{\const}{\text{const.}}
\DeclareMathAlphabet{\matheub}{U}{eur}{m}{n}
\newcommand{\termA}{\matheub{A}}
\newcommand{\termB}{\matheub{B}}
\newcommand{\termC}{\matheub{C}}
\newcommand{\termD}{\matheub{D}}

\newcommand{\termX}{\matheub{X}}
\newcommand{\PhiTilde}{\tilde{\Phi}}

\newcommand{\Aset}{\gpset{A}}
\newcommand{\indicator}[1]{\textrm{I}_{#1}}

\begin{document}
\title{The Edwards-Wilkinson equation with drift and Neumann boundary conditions}
\author{Seng Cheang and Gunnar Pruessner}
\ead{g.pruessner@imperial.ac.uk, seng.cheang@imperial.ac.uk}
\address{Department of Mathematics,
Imperial College London,
180 Queen's Gate,
London SW7~2AZ,
UK}
\date{5 May, 2010}

\begin{abstract}
The well known scaling of the Edwards-Wilkinson equation is essentially
determined by dimensional analysis. Once a drift term is added, more
sophisticated reasoning is required, which initially suggests that the drift term
dominates over the diffusion. However, the diffusion term is dangerously
irrelevant and the resulting scaling in fact non-trivial. In the present
article we compare the resulting scaling of the Edwards-Wilkinson
equation with drift and Neumann boundary
conditions to the published case with Dirichlet boundary conditions.
\end{abstract}

\submitto{\JPA}


\maketitle

\section{Introduction}
The Edwards-Wilkinson equation \cite{EdwardsWilkinson:1982} is very well
understood. It is a stochastic equation of motion describing the most
basic surface evolution of a growth model, consisting merely of
diffusion and random particle deposition. Given its minimal
parameterisation in terms of a diffusion constant $D$, a noise amplitude
$\Gamma$ as well as the system size $L$ and time $t$, 
\Eref{EW_eq} and \eref{physical_noise},
the scaling of its
roughness in $L$
and $t$ is
determined by dimensional analysis, because the only dimensionless
quantity is $Dt/L^2$.
This changes, as soon as another
interaction term is added whose coupling allows an additional
dimensionless quantity to be formed. In principle, the scaling is
then not determined by simple dimensional arguments. In this situation, it is common to invoke physical arguments 
to show that one of the terms is (infrared) irrelevant and thus can be
dropped
\cite{BarabasiStanley:1995,Taeuber:2005,LeBellac:1991}. There is no mathematically rigorous argument
underpinning this procedure and it therefore can produce incorrect
results. This has been analysed in the past
\cite{Pruessner_growthdrift:2004} for the
addition of a drift (convection) term to the Edwards-Wilkinson equation, which first
seems to render the diffusion irrelevant, when in fact it becomes
\emph{dangerously irrelevant} \cite{Ma:1976} and thus should not be dropped
from the analysis.

It has been noted \cite{Pruessner_growthdrift:2004} that the r{\^o}le of the drift term
depends on the boundary condition. It is trivial to show that in the
presence of periodic boundaries the drift does not enter the roughness
(in the definition below) at all, but changes the scaling when Dirichlet
boundary conditions are applied. One might now wonder to what extend the
result is universal, \ie what happens, for example, if Neumann boundary conditions are
applied. The calculations are far more involved than for the Dirichlet
case and presented in the following. 

It is not surprising if boundary conditions change some universal
quantities at the critical point \cite{PrivmanHohenbergAharony:1991},
certainly not if there is some net transport across the system as in the
present case. Given that the system is apparently sensitive to the
choice of boundary condition, the finding below might come as a
surprise; Neumann and Dirichlet boundary conditions produce, in
fact, the same leading order behaviour.

Beyond its r{\^o}le of probing scaling and universality in the
Edwards-Wilkinson equation,
the drift term has a simple physical motivation. It could, for
example, be caused by a gravitational or an electric field, present during
epitaxial growth. The choice of
the form of the boundary condition is a bit more arbitrary. A
Dirichlet boundary condition corresponds to particles falling off the
edge of the substrate. At the same time, particles are to be generated
whenever a hole forms at the boundary, which makes the boundary
condition somewhat unphysical. As discussed below, the Neumann
boundary is an improvement in this respect, but still does not
correspond an easily implemented physical situation, for example
by allowing no net flux across the boundary. The motivation for the
choice of the Neumann boundary condition in the following is thus the
question to what extent the choice of the boundary condition changes the
universal behaviour of the system compared to the Dirichlet case.

\subsection{Preliminaries}
The features of 
the original Edwards-Wilkinson (EW) equation \cite{EdwardsWilkinson:1982}
\begin{equation}
\partial_t \phi(\xvec,t) = D \nabla^2 \phi(\xvec,t) + \eta(\xvec,t)
\elabel{EW_eq}
\end{equation}
have been reviewed a number of times
\cite{BarabasiStanley:1995,Krug:1997,Meakin:1998,Vicsek:1999,Godreche:1992}. 
The field $\phi(\xvec,t)$ describes the height of an
interface over a substrate of dimension $d$ and linear extension $L$ at position $\xvec$ and time
$t$. The noise $\eta(\xvec,t)$ is Gaussian and white, characterised by
the usual correlator
\begin{equation}
\ave{\eta(\xvec,t)\eta(\xvec',t')} = 
\Gamma^2 \delta(\xvec-\xvec') \delta(t-t')
\elabel{physical_noise}
\end{equation}
with amplitude $\Gamma^2$ and $\ave{\eta}=0$. 
The operation
$\ave{\cdot}$ describes the averaging over the noise, such that, for
example, $\partial_t \ave{\phi(\xvec,t)} = D \nabla^2
\ave{\phi(\xvec,t)}$.
In the context of growth phenomena, the observable of choice
is often the roughness, which is the ensemble average of the spatial
variance of the field $\phi(\xvec,t)$
\begin{equation}
w^2(L,t) = \ave{ \spave{\phi(\xvec,t)^2} - \spave{\phi(\xvec,t)}^2}
\elabel{def_w2}
\end{equation}
where $\spave{\cdot}$ describes the spatial average
\begin{equation}
\spave{\cdot} = \frac{1}{L^d} \int_{L^d} \ddint{x} \cdot
\end{equation}
over the entire substrate of volume $L^d$. The roughness is closely
related to the correlation function
$C(\xvec,\xvec',t)=\ave{\left(\phi(\xvec,t)-\phi(\xvec',t)\right)^2}$, as
\begin{equation}
w^2(L,t) = \frac{1}{2} \frac{1}{L^d} \int_{L^d} \ddint{x'} \frac{1}{L^d} \int_{L^d}
\ddint{x} C(\xvec,\xvec',t)  \ .
\end{equation}

The scaling of the roughness in time is characterised by the universal
dynamical exponent $\dynamicalExpo$ and the finite size scaling in $L$ by the
universal roughness exponent $\roughnessExpo$, which is summarised in
the Family-Vicsek scaling hypothesis \cite{FamilyVicsek:1985}
\begin{equation}
w^2(L,t) = a L^{2\roughnessExpo} 
\GC\left(\frac{t}{b L^{\dynamicalExpo}} \right)
\elabel{FamilyVicsek_scaling}
\end{equation}
where $a$ and $b$ are two non-universal metric factors
\cite{PrivmanHohenbergAharony:1991}. The (universal) dimensionless scaling function $\GC(x)$ mediates between
two regimes, as $\lim_{x\to\infty} \GC(x)=\GC_\infty$ with $0<\GC_\infty<\infty$ and
$\GC(x)\sim x^{2\rougheningExpo}$ for $x\to0$ with the
roughening exponent $\rougheningExpo$ obeying
$\dynamicalExpo\rougheningExpo=\roughnessExpo$. 
Taking the limit $t\to\infty$ first, the roughness thus
scales like $w^2(L,t) \propto L^{2\roughnessExpo}$ in increasing $L$;
taking the thermodynamic limit $L\to\infty$ first,
it scales like $w^2(L,t)\propto t^{2\rougheningExpo}$ with increasing
$t$. Subleading terms in $L$ and $t$ respectively are expected to
be suppressed with increasing $t$ and $L$ respectively.

Given that the Edwards-Wilkinson equation in the form \eref{EW_eq} is
fully parameterised by $L,t$ and $D,\Gamma^2$, the \emph{only} possible
scaling is 
\begin{equation}
w^2(L,t;D,\Gamma^2) = \frac{\Gamma^2}{D} L^{2-d} \GC\left(\frac{Dt}{L^2}\right)
\elabel{EW_dim_ana}
\end{equation}
so that $\roughnessExpo=(2-d)/2$ and $\dynamicalExpo=2$, unless $w^2$
does not exist, diverges or
vanishes. In fact, at and above the upper critical dimension $d_c=2$,
the roughness is controlled by a lower cutoff or lattice spacing $a$,
and diverges for $d>d_c$ like $a^{2-d}$ as $a\to0$ \cite{Krug:1997}.

In the following, the Edwards-Wilkinson equation is studied in $d=1$
dimensions, where
\begin{equation}
\roughnessExpo=1/2 \quad \rougheningExpo=1/4 \quad \dynamicalExpo=2 
\qquad \text{(standard EW)} \ .
\end{equation}
Provided that no other scales are present, the result
\eref{EW_dim_ana} from dimensional analysis determines the scaling. Boundary
conditions generally affect the scaling function $\GC(x)$, and the
metric factors $a$ and $b$, \eref{FamilyVicsek_scaling}, but not the
exponents. Some types of interactions can be added to the
Edwards-Wilkinson equation without affecting its scaling behaviour, which
hints at the universality alluded to earlier. This can be illustrated by
adding a term $-\nu (\nabla^2)^2\phi$ on the right of \Eref{EW_eq}, which
results in a scaling of the roughness like
\begin{equation}
w^2(L,t;D,\Gamma^2,\nu) = \frac{\Gamma^2}{D} L^{2-d}
\GC'\left(\frac{Dt}{L^2},\frac{t \nu}{L^4}\right) \ .
\end{equation}
Irrespective of $t$,
the parameter $t \nu/L^4$ vanishes with increasing $L$ much faster than
the parameter $t D/L^2$, so that $w^2(L,t;D,\Gamma^2,\nu)$ approaches
$w^2(L,t;D,\Gamma^2,0)$, assuming $\GC'(Dt/L^2,t\nu/L^4)\approx\GC(Dt/L^2)$
for sufficiently large $L$ and irrespective of $t$, in particular
irrespective of whether $t$ is held fixed or the limit $t\to\infty$ is
taken. 
This
amounts to the statement that $-\nu (\nabla^2)^2\phi$ is asymptotically
irrelevant, so that the scaling of the Edwards-Wilkinson equation
of the original form \eref{EW_eq} is recovered.

\section{The Edwards Wilkinson equation with drift}
Whether a term is deemed relevant is a matter of the canonical dimension
of the coupling. Changing \eref{EW_eq} to 
the one-dimensional Edwards-Wilkinson equation with drift (EWd), 
\begin{equation}
\partial_t \phi(x,t) = D \partial_x^2 \phi(x,t) + v \partial_x \phi(x,t)
+ \eta(x,t) \ ,
\elabel{EWd_eq}
\end{equation}
at first seems to suggest that $D$ is irrelevant at any finite $v$,
suggesting a scaling of the form
\begin{equation}
w^2(L,t;v,\Gamma^2) = \frac{\Gamma^2}{v} L^{1-d}
\GC''\left(\frac{tv}{L}\right) \ ,
\end{equation}
so that
\begin{equation}
\roughnessExpo=0 \quad \rougheningExpo=0 \quad \dynamicalExpo=1 
\qquad \text{(suspected EWd)} \ .
\elabel{EWd_suspected_expos}
\end{equation}
This result, however, is obviously flawed, if periodic boundary
conditions (PBC) are applied. In that case the drift term can be
transformed away by a Galilean transformation, $\phitilde(x,t)=\phi(x-vt,t)$,
so that $\phitilde(x,t)$ follows the the original EW equation
\eref{EW_eq} and thus the roughness of the interface displays the
scaling derived in \eref{EW_dim_ana},
\begin{equation}
\roughnessExpo=1/2 \quad \rougheningExpo=1/4 \quad \dynamicalExpo=2
\qquad \text{(EWd with PBC)} \ .
\elabel{EWd_PBC}
\end{equation}
Since the initial guess
\eref{EWd_suspected_expos} is based on a purely physical argument
(rather than a mathematical one), this result merely questions the
validity of this type of reasoning. It is clear that if $v$ does not
actually enter into the observable $w^2$ as defined in \eref{def_w2} 
because it can be transformed away, then it cannot dominate its scaling.
As done below, where $v$ can \emph{not} be transformed away, one might
equally argue that $D$ never becomes truly irrelevant, 
\ie the exponents \eref{EWd_PBC} are a result of $D$ being dangerously
irrelevant \cite{Ma:1976}.

\latin{A priori} nothing can thus be said about the relevance of the
couplings $D$ and $v$. The scaling of the roughness has to be written as
\begin{equation}
w^2(L,t;D,\Gamma^2,v) = \frac{\Gamma^2}{D} L 
\GCtilde\left(\frac{Dt}{L^2},\frac{vt}{L}\right)
\end{equation}
which no longer fixes the scaling exponents $\roughnessExpo$,
$\rougheningExpo$ and $\dynamicalExpo$, as defined through
\Eref{FamilyVicsek_scaling}, based on a scaling function on a
single argument. The problem is the appearance of the new dimensionless
parameter $vt/L$, which can alter the asymptotic behaviour of $w^2$ in a
completely unknown way. In fact, with Dirichlet boundary conditions
(BC), $\phi(x=0,t)=\phi(x=L,t)=0$, it
was found \cite{Pruessner_growthdrift:2004} that $\GCtilde(Dt/L^2,vt/L)$ scales like
$(D/(vL))^{1/2}$ as $t\to\infty$ at fixed $L$ and like 
$(Dt/L^2)^{1/2}$ for $L\to\infty$ at fixed $t$, in summary
\begin{equation}
w^2(L,t;D,\Gamma^2,v) = \Gamma^2 \sqrt{\frac{L}{Dv}}
\GCtilde'\left(\frac{vt}{L}\right) \ ,
\end{equation}
with $\GCtilde'(x)\to\const$ as $x\to0$ and $\GCtilde'(x)\propto
\sqrt{x}$ as $x\to\infty$ (with corrections in powers of $Dt/L^2$),
so that
\begin{equation}
\roughnessExpo=1/4 \quad \rougheningExpo=1/4 \quad \dynamicalExpo=1
\elabel{EWd_DBC}
\qquad \text{(EWd with Dirichlet BC)} \ .
\end{equation}

Depending on the boundary condition, the additional scale $v$ thus
either leaves the scaling of the Edwards-Wilkinson equation unchanged
(as seen in the case of periodic BC), or changes them to anomalous
values, which cannot be recovered from a simple dimensional analysis (as
in the case of Dirichlet BC). The remainder of the present article is
dedicated to the question which scaling behaviour is generated by Neumann boundary conditions, $\partial_x\phi(x=0,t)=\partial_x\phi(x=L,t)=0$, 
which could lead, in principle, to a third set of exponents.

\subsection{Neumann boundary conditions}
If $\phi(\xvec,t)$ is the height of a grown surface at time $t$ atop a substrate at
position $\xvec$, the Dirichlet boundary condition mentioned above seems
slightly unphysical, because it is difficult to imagine a mechanism that
would pin the height to a certain value at the boundary of the surface,
by taking up any excess amount of matter and providing it when needed.
A zero flux boundary condition is much more naturally implemented, as it
corresponds to
imposing that no matter can enter or leave the substrate at the
boundaries. The flux in \eref{EWd_eq} is 
\begin{equation}
j(x,t) = - \left( D \partial_x + v \right) \phi(x,t)
\end{equation}
such that $\partial_t \phi = - \partial_x j$ as a matter
of mass balance (ignoring the noise). The condition $j(x,t) = 0$ at
$x=0,L$ is, however, very difficult
to analyse. Since the true motivation of this study is the question to what
extent the Edwards-Wilkinson equation with drift displays universal
behaviour, in the following the weaker Neumann condition
$\partial_x \phi (x,t) =0$ for $x=0,L$ will be analysed. 

To ease notation, it is helpful to absorb the various couplings and
other dimensionful quantities in 
\eref{EWd_eq} into a redefinition of time, space, field and noise. With
$y=x/L\in[0,1]$, 
$\tau=Dt/L^2$ and coupling $q=vL/D$ \Eref{EWd_eq} becomes
\begin{equation}
\partial_\tau \varphi(y,\tau) = \partial_y^2 \varphi(y,\tau) + q
\varphi(y,\tau) + \xi(y,\tau)
\elabel{EWd_eq_dimless}
\end{equation}
where
\begin{equation}
\elabel{def_varphi}
\varphi(y,\tau) = \sqrt{\frac{D}{L \Gamma^2}} \phi(x,t)
\text{ and } 
\xi(y,\tau) = \sqrt{\frac{L^3}{D \Gamma^2}} \eta(x,t) 
\end{equation}
so that
\begin{equation}
\ave{\xi(y,\tau)\xi(y',\tau')} = \delta(y-y')
\delta(\tau-\tau') \ .
\elabel{dimless_noise}
\end{equation}
The Neumann boundary conditions correspond to
\begin{equation}
\left.\partial_y\right|_{y=0,1} \varphi(y,\tau)=0 \ .
\end{equation}
As suggested by naive dimensional analysis, the coupling $q$ seems to
play an ever increasingly important r{\^o}le, as it diverges in the
thermodynamic limit $L\to\infty$. An alternative re-parameterisation of
the drift that
does not suffer from this problem is the dimensionless quantity 
$u=tv^2/D$, which will be of great use below.

The formal solution of \Eref{EWd_eq_dimless} for a given realisation of the
noise $\xi(y,\tau)$ 
\begin{equation}
\varphi(y,\tau) = \int_0^{\tau} \dint{\tau'} \int_0^1\dint{y'}
G(y,\tau-\tau';y',q)
\xi(y',\tau')
\elabel{formal_solution}
\end{equation}
is based on the Green function 
$G(y,\tau-\tau';y',q)$ 
which describes the
propagation of a Dirac delta peak at $y'$ at time $\tau'$ to $y$ at time
$\tau$. It is determined by considering the deterministic, homogeneous
partial differential equation  (PDE)
\begin{equation}
\partial_{\tau} G = (\partial_y^2 + q
\partial_y) G
\elabel{pde_G}
\end{equation}
 with initial condition
$\lim_{\tau\to0}G(y,\tau;y_0,q)=\delta(y-y_0)$ 
and Neumann boundary
conditions, which in turn can be constructed from a complete
set of eigenfunctions of the Sturm-Liouville
problem 
\begin{equation}
\lambda_n g_n(y) = (\partial_y^2 + q \partial_y) g_n(y) \ .
\end{equation}
The operator can be made self-adjoint with the help of a suitable weight
function, as discussed in great detail in \cite{FarkasFulop:2001},
$\braket{f}{g} = \int_0^1 \dint{y} e^{qy} f(y) g(y)$. The set of
normalised eigenfunctions orthogonal with respect to this scalar product
is then found as
\begin{equation}
g_n(y) =  e^{-\half q y} \sqrt{\frac{2}{k_n^2 + \quarter q^2}}
\left[k_n \cos(k_n y) + \half q \sin(k_n y) \right]
\end{equation}
and $\lambda_n=-\quarter q^2 - k_n^2$ for $n\ge1$, where $k_n=\pi n$.
The eigenfunction without a node, $n=0$, deviates from that pattern,
\begin{equation}
g_0(y) = \sqrt{\frac{q}{e^q-1}}
\end{equation}
and $\lambda_0=0$. Having introduced a scalar product above that renders
the differential operator self-adjoint, the temporal evolution of any
initial distribution under the PDE \eref{pde_G} can be determined.
On this basis, or equivalently, on the basis of completeness, the Green
function is found to be \cite[p. 63]{Butkovskiy:1982}
\begin{eqnarray}
\fl
 G(y,\tau;y_0,q)  =   e^{q y_0} \frac{q}{e^q-1}    
                 + e^{-\half q (y-y_0) - \quarter q^2 \tau} 
\sum_{n=1}^\infty \frac{2 e^{-k_n^2 \tau}}{k_n^2 + \quarter q^2} \nonumber\\
 \times \left(k_n \cos(k_n y_0) + \half q \sin(k_n y_0) \right) 
 \left(k_n \cos(k_n y)   + \half q \sin(k_n y) \right)
\elabel{full_propagator_kspace}
\end{eqnarray}

\subsubsection{Poisson summation}
In principle, the Green function \eref{full_propagator_kspace} could now
be used in \eref{formal_solution} and the asymptotic properties of the
roughness \eref{def_w2} be determined. However, it very quickly turns
out that the real space limit of large $L$ are very difficult to handle
in Fourier space, $k_n$, and the sums appearing in
\eref{full_propagator_kspace} are practically intractable. 
A Poisson summation could be performed at any stage, but is most easily
done directly on 
$G(y,\tau;y_0,q)$ 
itself. To this end, one notes that
\begin{eqnarray}
\fl
\frac{1}{k_n^2 + \quarter q^2}
\left(k_n \cos(k_n y_0) + \half q \sin(k_n y_0) \right)
\left(k_n \cos(k_n y)   + \half q \sin(k_n y) \right)
=  \nonumber \\
  \half \Big( \cos(k_n (y-y_0)) + \cos(k_n (y+y_0)) \Big) \nonumber \\
+ \half \frac{q}{k_n^2 + \quarter q^2}
        \left(k_n \sin(k_n (y+y_0)) - \half q \cos(k_n (y+y_0)) \right)
	\ ,
\end{eqnarray}
so that
\begin{eqnarray}\elabel{full_propagator_kspace_rewritten}
\fl 
G(y,\tau;y_0,q)  = e^{q y_0} \frac{q}{e^q-1}
 +   e^{-\half q (y-y_0) - \quarter q^2 \tau}  \nonumber \\
\times
        \sum_{n=1}^\infty e^{-k_n^2 \tau} 
          \Big(\cos(k_n (y-y_0)) + \cos(k_n (y+y_0)) \Big) \nonumber \\
 +  e^{-\half q (y-y_0) - \quarter q^2 \tau}
        \sum_{n=1}^\infty \frac{q e^{-k_n^2 \tau}}{k_n^2 + \quarter q^2} \nonumber\\
\times
          \left(k_n \sin(k_n (y+y_0)) - \half q \cos(k_n (y+y_0)) \right) \ .
\end{eqnarray}
This might not look like a significant improvement unless one notes that
\cite[p. 373]{MagnusOberhettingerSoni:1966}
\begin{equation}
\theta_3(z,\tau) =  1 + 2 \sum_{n=1}^\infty e^{\imag \pi \tau n^2} \cos(2 n \pi z) 
                 = \frac{1}{\sqrt{-\imag \tau}}
		       \sum_{n=-\infty}^\infty e^{-\imag \frac{\pi (z+n)^2}{\tau}}
\end{equation}
which allows the re-summation of the two cosine terms in
\eref{full_propagator_kspace_rewritten}, 
\begin{eqnarray}
\elabel{full_propagator_kspace_rewritten_again} 
\fl
G(y,\tau;y_0,q) = e^{q y_0} \frac{q}{e^q-1} - e^{-\half q (y-y_0) - \quarter q^2 \tau} 
               + e^{-\half q (y-y_0) - \quarter q^2 \tau}
                      \sum_{n=1}^\infty \frac{q e^{-k_n^2 \tau}}{k_n^2 +
		      \quarter q^2} \nonumber \\
    \times	      \left(k_n \sin(k_n (y+y_0)) - \half q \cos(k_n (y+y_0)) \right) 
               + e^{-\half q (y-y_0) - \quarter q^2 \tau}
  \frac{1}{\sqrt{4\pi\tau}} \nonumber \\
\times \sum_{n=-\infty}^\infty
  \left(e^{-\frac{(y-y_0+2n)^2}{4\tau}} + e^{-\frac{(y+y_0+2n)^2}{4\tau}}
  \right) 
\ . 
\elabel{full_propagator_realspace}
\end{eqnarray}
Without the exponential prefactor, $\exp[-\half q (y-y_0) - \quarter q^2
\tau]$, the last summation describes the diffusive motion of particles
with (positive) mirror image on a ring of circumference $2$ not being subject
to the drift. In the following, the aim is to express \Eref{full_propagator_realspace} in terms
of the free propagator on $\Rset$
\begin{equation}
\Phi_0(y,\tau;q) = \frac{1}{\sqrt{4\pi\tau}}
e^{-\frac{(y+q\tau)^2}{4\tau}} \ ,
\elabel{def_Phi0}
\end{equation}
which solves \eref{pde_G}, so that none of the terms is expressed any longer
in Fourier space, which facilitates
integration and the determination of the asymptote using a saddle
point approximation. In fact, including all pre-factors, 
the last sum in \eref{full_propagator_kspace_rewritten_again} can be written as

\begin{equation}
\sum_{n=-\infty}^\infty e^{qn}
  \Big(\Phi_0(y-y_0+2n, \tau;q) + e^{q y_0} \Phi_0(y+y_0+2n,
  \tau;q)\Big) \ .
\end{equation}

The first three terms in
\eref{full_propagator_kspace_rewritten_again}, the two
exponentials and the sum, are much more difficult to handle. Taking the limit $\tau \rightarrow 0$, the expression

\begin{eqnarray}
\fl
s(y,y_0)  =   e^{q y_0} \frac{q}{e^q-1} - e^{-\half q (y-y_0)} 
          + e^{-\half q (y-y_0)} \sum_{n=1}^\infty \frac{q}{k_n^2 +\quarter q^2} \nonumber \\
          \times   \left(k_n \sin(k_n (y+y_0)) - \half q \cos(k_n (y+y_0)) \right) \\
         =  \frac{q e^{q y_0}}{e^q-1} + \half q e^{-\half q (y-y_0)}
\sum_{n=-\infty}^\infty \frac{1}{k_n^2 +\quarter q^2} \nonumber\\
\times \left(k_n \sin(k_n (y+y_0)) - \half q \cos(k_n (y+y_0)) \right)
\end{eqnarray}
can be regarded as an initial source which is propagated under the PDE
\eref{pde_G} and on $\Rset$ by the propagator \eref{def_Phi0}, so that
in fact
\begin{eqnarray}
\fl
G(y,\tau;y_0,q)  = \int_{-\infty}^{\infty} \dint{\ytilde}
s(\ytilde,y_0) \Phi_0(y-\ytilde, \tau;q) \nonumber \\
 +\sum_{n=-\infty}^\infty e^{qn}
  \Big(\Phi_0(y-y_0+2n, \tau;q) + e^{q y_0} \Phi_0(y+y_0+2n, \tau;q)\Big)
  \ .
\end{eqnarray}
Taking the limit $\tau\to0$ for the entire 
$G(y,\tau;y_0,q)$,
\eref{full_propagator_kspace_rewritten_again} reveals
\begin{equation}
\fl
\lim_{\tau\to0} G(y,\tau;y_0,q) 
=  s(y,y_0) + e^{-\half q (y-y_0)}  \sum_{n=-\infty}^\infty \left( \delta(y-y_0+2n) + \delta(y+y_0+2n) \right)
\end{equation}
which means that $s(y,y_0)=0$ for $y,y_0\in[0,1]$, since 
$\lim_{\tau\to0}G(y,\tau;y_0,q)=\delta(y-y_0)$ 
on that interval, which implies that
\begin{equation}
\sum_{n=-\infty}^\infty \frac{1}{k_n^2 +\quarter q^2}
\left(k_n \sin(k_n z) - \half q \cos(k_n z) \right)
=
-\frac{2}{e^q-1} e^{\half q z}
\end{equation}
for any $z\in[0,2]$. Clearly, the left hand side is periodic in $z$ with
period $2$, which means that it is in fact the periodic repetition of
the right hand side, \ie
\begin{equation}
\fl
\sum_{n=-\infty}^\infty \frac{1}{k_n^2 +\quarter q^2}
\left(k_n \sin(k_n z) - \half q \cos(k_n z) \right)
=
- \sum_{n=-\infty}^\infty \rho(z+2n;q)
\end{equation}
with 
\begin{equation}
\rho(z;q) = \frac{2}{e^q-1} e^{\half q z} I_{[0,2]}(z)
\end{equation}
and the indicator function
\begin{equation}
\elabel{def_I}
I_{\Omega}(z) = \left\{
\begin{array}{ll}
1 & \text{ for } z\in\Omega\\
0 & \text{ otherwise \ .}
\end{array}
\right.
\end{equation}
On this basis, the source can be rewritten as
\begin{equation}
s(y,y_0) = \frac{q e^{q y_0}}{e^q-1}
\left(1 - \sum_{n=-\infty}^\infty e^{qn} I_{0,2}(y+y_0+2n)\right)
\end{equation}
and therefore
\begin{eqnarray}
\elabel{realspace_propagator}
\fl
G(y,\tau;y_0,q) =
\int_0^2 d\ytilde \frac{q e^{q y_0}}{e^q-1} \sum_{n=-\infty}^\infty (1-e^{qn})
\Phi_0(y+y_0+2n-\ytilde, \tau;q) \nonumber \\
        +  \sum_{n=-\infty}^\infty e^{qn}
  \Big(\Phi_0(y-y_0+2n, \tau;q) + e^{q y_0} \Phi_0(y+y_0+2n, \tau;q)\Big)
\ .
\end{eqnarray}
Problems of convergence have not been addressed here in a detail, which
could affect some of the manipulations done above, in particular when
the order of integration and summation is changed. Yet, because of the
free propagator $\Phi_0$ effectively posing an exponential cutoff on the
sums as well as the integrals, all convergence issues turn out to be
harmless.

To verify that the propagator \eref{realspace_propagator} indeed solves
the PDE \eref{pde_G} is a matter of a straight forward calculation.
Similarly, the
initial condition 
$\lim_{\tau\to0} G(y,\tau;y_0,q)=\delta(y-y_0)$ for $y,y_p\in[0,1]$
can be identified by mere inspection. The only slight difficulty are
the boundary conditions 
$\partial_y|_{y=0,1} G(y,\tau;y_0,q)=0$, which are
most easily verified by writing the gradient as
\begin{eqnarray}
\fl
\partial_y G(y,\tau;y_0,q)  =
e^{-\half q (y-y_0)-\quarter q^2 \tau} \frac{1}{\sqrt{4 \pi \tau}} 
\Bigg\{
\sum_{n=-\infty}^\infty \frac{y_0-2n-q\tau}{2\tau} 
     \left(e^{-\frac{(y-y_0+2n)^2}{4\tau}}-e^{-\frac{(-y-y_0+2n)^2}{4\tau}}\right) \nonumber \\
  -\frac{y}{2\tau} 
     \left(e^{-\frac{(y-y_0+2n)^2}{4\tau}}+e^{-\frac{(-y-y_0+2n)^2}{4\tau}}\right)
\Bigg\}
\end{eqnarray}

\subsection{Calculation of the roughness}
The propagator \eref{realspace_propagator} can be used in the formal solution
\eref{formal_solution}, which provides the basis for calculating the
roughness via \eref{def_w2} and \eref{def_varphi}, 
\begin{equation}
\elabel{roughness_with_prefac}
w^2(L,t) = \frac{L \Gamma^2}{D} 
\int_0^1\dint{y_1}
\int_0^1\dint{y_2}
\big[\delta(y_1-y_2)-1\big]
\varphi(y_1,\tau)\varphi(y_2,\tau) \ ,
\end{equation}
where the ensemble average enters only through \eref{dimless_noise}.

The integral on the right is the dimensionless roughness, 
\begin{equation}
\elabel{dimless_roughness_first_attempt}
\int_0^1\dint{y_1}
\int_0^1\dint{y_2}
\big[\delta(y_1-y_2)-1\big]
\varphi(y_1,\tau)\varphi(y_2,\tau) \ ,
\end{equation}
which depends on only two dimensionless parameters, $\tau=Dt/L^2$ and
$q=vL/D$.
However, as discussed in the following, the parameterisation
\begin{equation}
\wtilde^2(q,u) = q 
\int_0^1\dint{y_1}
\int_0^1\dint{y_2}
\big[\delta(y_1-y_2)-1\big]
\varphi(y_1,\tau)\varphi(y_2,\tau) \ ,
\end{equation}
with $u=tv^2/D=q^2\tau$ is advantageous, so that $w^2(L,t)=(\Gamma^2/v)
\wtilde^2(q,u)$, which means that all scaling of the roughness can be
read off the scaling of $\wtilde^2$ without any further prefactor such
as $L$ in \eref{roughness_with_prefac}.

The limit $t\to\infty$ in \eref{roughness_with_prefac} translates simply to $\tau\to\infty$ 
in \eref{dimless_roughness_first_attempt} and the
scaling of this limit in $L$ is observed through the scaling in $q$, which
is unaffected by the limit $\tau\to\infty$. So, in
order to determine the roughness exponent $\roughnessExpo$, the
parameterisation of \eref{dimless_roughness_first_attempt} in $q,\tau$ is suitable. However, taking the
thermodynamic limit $L\to\infty$ means $\tau\to0$ and $q\to\infty$
simultaneously and not independently, since $q^2\tau=u$ remains constant.

Using $\wtilde^2(q=vL/D,u=tv^2/D)$ as the parameterisation of the roughness means
that the two limits $L\to\infty$ and $t\to\infty$ affect each only one
dimensionless parameter but not the other. The limit $t\to\infty$ means
$u\to\infty$ and the scaling in large $L$ is observed through
arbitrarily large but fixed $q$. 
The limit $L\to\infty$ means $q\to\infty$ and the scaling in large $t$
is observed through the scaling in large but fixed $u$. The two
different asymptotes will be determined by considering large $q$ in
both cases and $u/q$ diverging or $u/q$ vanishing.

Using the parameter $u$ in favour of $\tau$ needs to be implemented in
the propagator as well, \eref{realspace_propagator}, using $\tau=u/q^2$.
It is also useful to replace the integral over $\tau$ in the formal
solution \eref{formal_solution} correspondingly by an integral over $u$
so that
\begin{eqnarray}
\frac{v}{\Gamma^2} w^2(L,t) = 
\wtilde^2(q,u) = 
q^{-1} \int_0^1 \dint{y_1} \int_0^1 \dint{y_2}
\big[\delta(y_1-y_2)-1\big] \nonumber\\
\times
\int_0^1 \dint{y'}\ \int_0^u \dint{u'}
        \Gtilde(y_1,u';y',q)
	 \Gtilde(y_2,u';y',q) \ ,
\elabel{w2_integration}
\end{eqnarray}
where $\Gtilde$ is the re-parameterised propagator with all
parameters explicitly appearing as arguments.

The bulk of the work is performing the integration \eref{w2_integration}. This can be done more
conveniently by
splitting the propagator up into four terms, 
\begin{equation}
\Gtilde(y_i,u';y',q) = \termA_i - \termB_i + \termC_i + \termD_i \ ,
\elabel{def_propagator}
\end{equation}
where
\begin{subequations}
\elabel{all_terms}
\begin{eqnarray}
\termA_i & \equiv &
  \frac{q e^{q y'}}{e^q - 1} 
\elabel{termA}
\\
\termB_i & \equiv &
  \frac{q e^{q y'}}{e^q - 1}
  \sum_{n_i=-\infty}^\infty e^{qn_i} \int_0^2 \dint{\ytilde_i}
  \PhiTilde_0(y_i+y'+2n_i-\ytilde_i,u;q) \\
\termC_i & \equiv &
  \sum_{n_i=-\infty}^\infty e^{qn_i}
  \PhiTilde_0(y_i-y'+2n_i,u;q) \\
\termD_i & \equiv &
  \sum_{n_i=-\infty}^\infty e^{q(n_i+y')}
  \PhiTilde_0(y_i+y'+2n_i,u;q) \\
\end{eqnarray}
\end{subequations}
with
\begin{equation}
\PhiTilde_0(y,u;q)  = 
  \frac{q}{\sqrt{4\pi u}} 
  e^{-\frac{(qy+u)^2}{4u}} \ .
\elabel{def_PhiTilde}
\end{equation}

\subsection{Integration and analysis}
Because of the prefactor $\big[\delta(y_1-y_2)-1\big]$,
for each of the ten distinct terms generated by using \eref{all_terms} in
\eref{w2_integration}, effectively
two different integrations have to be performed, namely one over $y_1$
and $y_2$ distinct, and one over $y_1=y_2$. In addition, an integral over
$y'$ and $u'$ needs to be performed, as well as over $\ytilde$ for term
$\termB_i$. In all cases, a saddle point approximation (SPA) is used, specifically
\begin{equation}
\frac{q}{\sqrt{4\pi \utilde}}
\int_\Aset \dint{y} 
e^{-\frac{(q(y-y_0)+\utilde)^2}{4 \utilde}}
= 
 I_\Aset\left(y_0-\frac{\utilde}{q}\right) 
\quad \text{for }
 \utilde/q^2 \ll 1 
\ ,
\elabel{Gaussian_approx}
\end{equation}
where $I_{\Aset}(y)$ is again an indicator function, \eref{def_I}.
It is worth stressing that there are no further algebraic terms, \ie all
other contributions from the integral vanish exponentially in large
$q^2/\utilde$.

In most cases, further integrals over variables contained in the argument of
the indicator function are to be performed. If the intersection of the
integration range of the other variables and the set $\Aset$ above has
finite measure, the result is straight forward to calculate. If,
however, the intersection contains a single point, then the SPA has to be
considered to higher order. In general, for each ``marginal variable''
the resulting integral acquires an additional factor $\sqrt{u}/q$.
For example, to leading order
$\utilde/q^2$,
\begin{equation}
\int_0^1\dint{y_1} 
\int_0^1\dint{y_2}
e^{-\frac{(q(y_1-y_2-1/3))^2}{4u}} \approx
\frac{\sqrt{4\pi u}}{q} 
\int_0^1\dint{y_2} I_{(0,1)} (y_1+1/3)
=
\frac{4 \sqrt{\pi u}}{3q} 
\end{equation}
using \eref{Gaussian_approx} in a non-marginal case (maximum of the
exponential at $y_1-y_2=1/3\in(-1,1)$), whereas replacing $1/3$ by $1$
produces
\begin{equation}
\int_0^1\dint{y_1} 
\int_0^1\dint{y_2}
e^{-\frac{(q(y_1-y_2-1))^2}{4u}} \approx
\frac{2u}{q^2} 
\ ,
\end{equation}
rather than $\frac{\sqrt{4\pi u}}{q} \int_0^1\dint{y_2} I_{(0,1)}
(y_1+1)=0$.
Integrals like that result in subleading terms, whose amplitude is not
normally calculated in the following. Instead, only the power of $q$ is
noted and by comparison to other terms it is verified that it is safe to
ignore it. The same result is recovered by power counting in the four
terms \eref{all_terms}, where each integral gives rise to a leading
order $q^{-1}$ and
thus each of the four terms has the same algebraic dependence on $q$.

Further details of the calculation are exemplified in the appendix.
Combining all contributions gives
\begin{eqnarray*}
w^2_{\physical} &=& \frac{\Gamma^2}{v} 
q^{-1} \int_0^1 \dint{y_1} \int_0^1 \dint{y_2}
\big[\delta(y_1-y_2)-1\big]\\
&&\times
\left(
 \termA_1
-\termB_1
+\termC_1
+\termD_1
\right)
\left(
 \termA_2
-\termB_2
+\termC_2
+\termD_2
\right)
\end{eqnarray*}
For $u$ small, looking at the limit $q\rightarrow\infty$, to leading
order:
\begin{equation}
\lim_{q\rightarrow\infty} w^2_{\physical} = \frac{\Gamma^2}{v} \sqrt{\frac{u}{2\pi}}
\elabel{q_infty} 
\end{equation}
For $q$ large and considering the limit $u\rightarrow\infty$, to
leading order:
\begin{equation}
\lim_{u\rightarrow\infty} w^2_{\physical} 
= \frac{\Gamma^2}{v} \frac{2}{3\sqrt{2\pi}} \sqrt{q} 
\elabel{u_infty}
\end{equation}

\section{Discussion and conclusion}
Upon replacing $u$ and $q$ by their definitions, $u=tv^2/D$ and
$q=vL/D$, the two asymptotes for the roughness derived in \Eref{u_infty} and
\eref{q_infty} are
\begin{equation}
w^2(t,L) = \Gamma^2 \times 
\left\{
\begin{array}{rl}
\sqrt{\frac{t}{2\pi D}} & \textrm{for}\ L\to\infty \\
 & \\
\frac{2}{3} \sqrt{\frac{L}{2\pi D v}} & \textrm{for}\ t\to\infty\\
\end{array}
\right.
\elabel{qEW_final}
\end{equation}
which is, to leading order, identical to the result in
\cite{Pruessner_growthdrift:2004} for Dirichlet boundary conditions. 
The exponents
as defined in \eref{FamilyVicsek_scaling},
\begin{equation}
\alpha = \frac{1}{4},\ \ \ \beta = \frac{1}{4},\ \ \ z=1 
\elabel{EWd_NBC}
\qquad \text{(EWd with Neumann BC)}
\end{equation}
therefore reproduce \eref{EWd_DBC}. On the one hand, this is a
surprising result, because after realising that the usual dimensional
arguments do no longer apply, \emph{any} exponents are mathematically
possible. The fact that Neumann boundary conditions reproduce the
anomalous results of the Dirichlet case (contrasting those for periodic
boundary conditions) even down to the amplitudes 
therefore point to some universal mechanism at work in both equations.
Physically, on the other hand, arguments very similar to those discussed
in \cite{Pruessner_growthdrift:2004} apply: The drift with velocity $v$
effectively constantly re-initialises the interface with vanishing slope
from one side to the other, while constantly under the influence of the
external noise. This mechanism erases practically all features that develop over
times exceeding $L/v$ and thus in saturation, $t\to\infty$ reduces the roughness to
about the value after time $L/v$. In fact, the stationary roughness is $2/3$ of what is extrapolated
from initial roughening:
\begin{equation}
\lim_{t\to\infty} w^2(t,L) = 
\frac{2}{3} \lim_{L'\to\infty} w^2(L/v,L') 
\end{equation}

In conclusion we have shown that the Edwards-Wilkinson equation with
drift and Neumann boundary conditions produces anomalous scaling 
different from what is expected from naive dimensional analysis and
easily derived for periodic boundary conditions. 
The scaling and the amplitudes of the Dirichlet case are reproduced,
consistent with a simple physical scenario of an interface that is
constantly re-initialised.
Neumann boundary
conditions are physically more relevant than Dirichlet ones, yet far
more difficult to handle analytically.

\ack
GP would like to thank Vikram Pandya for bringing his
attention to this problem. We thank D. E. Khmelnitskii for pointing
out a mistake in an earlier version of this manuscript.

\appendix
\section{Integral \eref{w2_integration} term by term}
In this appendix some details of the integral \eref{w2_integration} are
shown, considering the propagator term by term as defined in
\eref{def_propagator} and \eref{all_terms}.

\subsection{$\termA_1 \termX_2$ Term}
Since there is no $y_1$ or $y_2$ dependence in $\termA_1$, \eref{termA}:
\begin{eqnarray*}
\int_0^1 \dint{y_1} 
\underbrace{\int_0^1 \dint{y_2} \big[\delta(y_1-y_2)-1\big]}_{\equiv 0}
\int_0^1 \dint{y'} \int_0^u \dint{u'} \termA_1 \termX_2 = 0
\end{eqnarray*}
for $\termX$ being any of the $\termA,\termB,\termC$ or $\termD$ terms.

\subsection{${\termB_1 \termX_i}$ Term}
Since all ${\termB_1\termX_i}$ integrals  follow 
similar arguments, we exemplify the procedure on ${\termB_1\termC_2}$. 
The integrals to be calculated (to leading order) are 
\begin{eqnarray}
\fl
w^2_{\termB\termC} = \frac{\Gamma^2}{v}q^{-1}
        \int_0^1 \dint{y_1}
                \int_0^1 \dint{y_2} \big[\delta(y_1-y_2)-1\big]
        \int_0^1 \dint{y'}
                \int_0^u \dint{u'} \elabel{BC_integral} \\
\fl
\times
  \frac{q e^{q y'}}{e^q - 1}
  \sum_{n_1=-\infty}^\infty e^{qn_1} \int_0^2 \dint{\ytilde_1}
\frac{q}{\sqrt{4\pi u}}
e^{-\frac{(q(y_1+y'+2n_1-\ytilde_1)+u)^2}{4u}}
  \sum_{n_2=-\infty}^\infty e^{qn_2}
\frac{q}{\sqrt{4\pi u}}
e^{-\frac{(q(y_2-y'+2n_2)+u)^2}{4u}} 
\nonumber \ .
\end{eqnarray}

We consider the total exponential of ${\termB_1\termC_2}$ in the form
\begin{equation}
\frac{1}{e^q-1} \exp{ \left(q(n_1 + n_2) - f(y',\ytilde) \right) } 
\elabel{exponent_b1c2}
\end{equation}
where
\begin{equation}
\fl
f(y',\ytilde) = -qy' + \frac{q^2}{4u'} \left(y_1+y'+2n_1-\ytilde+\frac{u'}{q} \right)^2 
   + \frac{q^2}{4u'} \left(y_2-y'+2n_2+\frac{u'}{q} \right)^2
\elabel{b1c2_f}
\ .
\end{equation}
A contribution in the form \eref{exponent_b1c2} vanishes for large $q$ unless
\begin{equation}
n_1+n_2 - \frac{f(y',\ytilde)}{q} \ge 1
\elabel{b1c2_con1}
\ .
\end{equation}
Applying the SPA to the integral over $y'$, means to evaluate the
integral at $y'$ that minimises $f$: 
\begin{equation}
y'_0 = \frac{1}{2} (y_2 - y_1) + (n_2 - n_1) + \frac{1}{2}\ytilde + \frac{u'}{q}
\ ,
\end{equation}
where $f$ now becomes:
\begin{equation}
f(y'_0,\ytilde) = \frac{q^2}{2u'} \left( \frac{1}{2}(y_1+y_2) + (n_1+n_2) - \frac{\ytilde}{2}  \right)^2 + q(y_1+2n_1-\ytilde)
\ .
\end{equation}
Applying the SPA again to the integral over $\ytilde$ gives, with
$f(y_0',\ytilde)$ above,
\begin{equation}
\ytilde_0 = y_1+y_2+2(n_1+n_2)+\frac{4u'}{q}
\ ,
\end{equation}
where now
\begin{equation}
f(y_0',\ytilde_0) = -q(y_2+2n_2) - 2u'
\elabel{b1c2_finalf}
\ .
\end{equation}
Determining the asymptotic behaviour is greatly facilitated by writing
$\frac{u'}{q} = \sigma + m\ge0$, with $\sigma \in [0,1)$ and $m\in
\Nset^0$, so that $\lfloor u'/q\rfloor=m$. 
Using the final result of $f$, \eref{b1c2_finalf}, the inequality in \eref{b1c2_con1} becomes:
\begin{equation}
n_1 + n_2 - \frac{f}{q} = y_2 +2\sigma + n_1 + 3n_2 + 2m \ge 1
\ .
\elabel{b1c2_con1_dash}
\end{equation}
As $y_2\in[0,1]$ and $\sigma \in [0,1)$, so $y_2 + 2\sigma \in [0,3)$, the
contribution \eref{exponent_b1c2} vanishes for large $q$ unless
\begin{equation*}
n_1+3n_2+2m > -2
\ .
\end{equation*}
Since $n_1, n_2$ and $m$ are integers, we have the inequality
\begin{equation}
n_1+3n_2+2m \geq -1
\ .
\elabel{b1c2_3}
\end{equation}
Both SPAs produce an indicator function as well. 
Writing $\frac{u'}{q}$ in terms of $\sigma$ and $m$ gives
\begin{eqnarray}
\indicator{(0,1)}(y_0') = \indicator{(0,1)} \left[ (y_2 - y_1)/2 +
(n_2-n_1) + \ytilde/2 + \sigma + m  \right]  
\elabel{b1c2_ind1a}  \\
\indicator{(0,2)}(\ytilde_0) = \indicator{(0,2)}\left[ y_1+y_2+2(n_1+n_2)+4\sigma + 4m \right]
\elabel{b1c2_ind2}
\ .
\end{eqnarray}
From the indicator function of $y_0'$ above, \eref{b1c2_ind1a}, we have
\begin{equation}
0 < \frac{1}{2} (y_2 - y_1) + (n_2 - n_1) + \frac{1}{2} \ytilde + \sigma + m <1
\elabel{b1c2_ind11a}
\end{equation}
as a condition for \eref{exponent_b1c2} to contribute. 
As $y_1,y_2 \in [0,1)$, $\ytilde \in [0,2)$ and $\sigma \in [0,1)$, this
implies that $\frac{1}{2} (y_2 - y_1) + \frac{1}{2} \ytilde + \sigma \in
[-\frac{1}{2},\frac{5}{2})$. By subtracting this range, \eref{b1c2_ind11a} becomes $-\frac{5}{2} < n_2 - n_1 + m < \frac{3}{2}$, which gives
\begin{equation}
-2 \leq n_2 - n_1 + m \leq 1
\elabel{b1c2_1}
\end{equation}
as $n_1,n_2$ and $m$ are integers.
Similarly, for \eref{b1c2_ind2} to be non-zero, we have the condition
\begin{equation}
-2 \leq (n_1+n_2) +2m\leq 0
\elabel{b1c2_2a}
\ .
\end{equation}

${\termB_1 \termC_2}$ vanishes if either one of these two ranges,
\eref{b1c2_1} and \eref{b1c2_2a}, is incompatible with all possible values of
$n_1,n_2$ and $m$. 
By adding \eref{b1c2_3} and \eref{b1c2_1} we find $-3\le4n_2+3m$ and as
$m\in\Nset^0$ it follows $n_2\ge0$.
Adding
\eref{b1c2_1} and \eref{b1c2_2a}, we obtain $-4\leq 2n_2 +
3m \leq 1$ and with $m\ge0$
it now follows $n_2=m=0$ to prevent
either indicator function from vanishing. Therefore, the two indicator
functions become
\begin{eqnarray}
\indicator{(0,1)}(y_0') = \indicator{(0,1)} [y_2+3\sigma] 
\elabel{b1c2_ind1aa}\\
\indicator{(0,2)}(\ytilde_0) = \indicator{(0,2)} [y_1+y_2+2n_1+4\sigma]
\elabel{b1c2_ind2aa} \ .
\end{eqnarray}
The next variable to determine is $n_1$. With $n_2=m=0$
\Eref{b1c2_con1_dash} reads:
\begin{equation}
y_2 + 2\sigma + n_1 > 1
\elabel{b1c2_con1_dash1}
\end{equation}
Multiplying \eref{b1c2_con1_dash1} by 2 and rearranging, we have the
condition $ y_2 + 4\sigma+2n_1 > 2-y_2 > 1$, as $y_2 \in [0,1)$,
and thus
\begin{equation}
y_2 + 3\sigma > 1-\sigma-2n_1
\elabel{b1c2_4}
\end{equation}
Using $n_2=m=0$ in \eref{b1c2_3} and \eref{b1c2_2a} gives the range of
$n_1$ as $-1 \leq n_1 \leq 0$, whereas \eref{b1c2_ind1aa} indicates
\begin{equation}
1 > y_2 +3\sigma > 0
\elabel{b1c2_ind1aaa}
\ .
\end{equation}
If $n_1=-1$, then \eref{b1c2_4} becomes $y_2+3\sigma > 3-\sigma$. Since
$\sigma \in [0,1)$, this contradicts \eref{b1c2_ind1aaa}. Therefore, we
conclude that $n_1=0$. However, if we use the fact that $n_1=n_2=m=0$ in
\eref{b1c2_con1_dash} we get $y_2+2\sigma > 1$. By comparing to
\eref{b1c2_ind1aaa}, we have $\sigma < 0$ which contradicts the fact
that $\sigma \in [0,1)$. As the conditions cannot be fulfilled
simultaneously, this implies that the integrals 
vanish exponentially in large $q$.

Considering in addition any marginal cases will produce terms of lower
algebraic order in $q$. For the present integral, the marginal cases are $n_1=n_2=m=0$ with $y_2=0$
and $\sigma=0$ (double marginal), as well as $n_1=1, n_2=m=0$ with
$y_1=y_2=0$ and $\sigma=0$  (triple marginal). Power counting in the
initial integral \eref{BC_integral} thus gives overall contributions to
$w^2_{\termB\termC}$ of order $\OC(q^{-2})$ and $\OC(q^{-3})$
respectively. Extra care must be taken when considering the integral
with upper bound $u$, as $u\to\infty$ might be taken before
$q\to\infty$, in which case the integral over $u'$ might give rise to a
term of order $q$ itself (see, for example, \eref{DD_qfirst} versus
\eref{DD_ufirst}). In the present case this does not apply,
because $u'=q(m+\sigma)$ and both $m$ as well as $\sigma$ are fixed by
the SPA.

By similar arguments, the terms $\termB_1\termB_2$ and $\termB_1\termD_2$ vanish.

\subsection{$\termC_1 \termX_2$ and $\termD_1 \termD_2$ Term} 
Calculations for the terms $\termC_1\termX_2$ and $\termD_1 \termD_2$ 
are very similar and we therefore exemplify the procedure for
$\termD_1\termD_2$ only.

We write contribution by $\termD_1\termD_2$ as
\begin{eqnarray*}
\fl
w^2_{\termD\termD} = \frac{\Gamma^2}{v}q^{-1}
        \int_0^1 \dint{y_1}
                \int_0^1 \dint{y_2} \big[\delta(y_1-y_2)-1\big]
        \int_0^1 \dint{y'}
                \int_0^u \dint{u'} \nonumber \\
\fl
\times
        \sum_{n_1=-\infty}^{\infty} e^{qn_1}
                                \frac{q}{\sqrt{4\pi u'}} e^{\frac{-[q(y_1+y'+2n_1)+u']^2}{4u'}} e^{qy'} 
\sum_{n_2=-\infty}^{\infty} e^{qn_2}
                \frac{q}{\sqrt{4\pi u'}} e^{\frac{-[q(y_2+y'+2n_2)+u']^2}{4u'}} e^{qy'}
\end{eqnarray*}

We consider the total exponent of $\termD_1\termD_2$ in the form
\begin{equation}
\exp(q(n_1+n_2)-f)
\end{equation}
We first apply the SPA to the integral over $y'$, and obtain the minimum $y'_{0}$,
so that
\begin{equation*}
\fl f(y'_{0}) = \frac{q^2}{2u'}
        \left[
                \frac{1}{2} (y_1-y_2) + (n_1-n_2)]
        \right]^2 + 2q \left[ \frac{1}{2} (y_1+y_2) + (n_1+n_2) \right]
\ .
\end{equation*}
Again, the analysis of the asymptotic behaviour is greatly facilitated
by writing $\frac{u'}{q} = \sigma + m$, with $\sigma \in [0,1)$ and
$m\in \Nset^0$, so that the indicator function for $y'_0$ is
\begin{equation}
\indicator{(0,1)}(y'_{0}) = \indicator{(0,1)} \left( \sigma - \frac{1}{2} (y_1+y_2) + m  - (n_1+n_2) \right)
\ .
\end{equation}
Since $\sigma - \frac{1}{2} (y_1 + y_2) \in [-1,1)$, 
the remainder of the argument $m-(n_1+n_2)$ must be 
greater than $-1$ or less than $2$ 
for this term to contribute. 
Therefore we will only need to consider only $n_1+n_2=m$ and
$n_1+n_2=m-1 $, incorporated below by means of Kronecker 
$\delta$-functions. The total contribution then becomes:
\begin{eqnarray*}
\fl
w^2_{\termD\termD} = \frac{\Gamma^2}{v}q^{-1}
        \int_0^1 \dint{y_1}
                \int_0^1 \dint{y_2} \big[\delta(y_1-y_2)-1\big] \\
 \times
                \int_0^u \dint{u'}
                        \sum_{n_1,n_2} 
                                \frac{q}{\sqrt{8\pi u'}} 
   \exp 
        \Bigg\{
           -\frac{q^2}{2u'}
           \left[
                \frac{1}{2} (y_1-y_2) + (n_1-n_2)
           \right]^2  \\
           - 2q \left[ \frac{1}{2} (y_1+y_2) + (n_1+n_2) \right] 
        \Bigg\}
 \Bigg\{
        e^{-qm} \indicator{(0,1)}\left[-\frac{1}{2} (y_1+y_2) + \sigma\right]\delta_{m,n_1+n_2}  \\
         +  e^{-q(1-m)} \indicator{(-1,0)}\left[-\frac{1}{2} (y_1+y_2) + \sigma\right]\delta_{m-1,n_1+n_2}
        \Bigg\}
\ .
\end{eqnarray*}
At this stage, it is sensible to consider the $\delta$-term and $1$-term
in $\big[\delta(y_1-y_2)-1\big]$ 
separately. For $y_1=y_2=y$, i.e. the contribution of the $\delta$-term,
the integral to consider is
\begin{eqnarray*}
\fl
(\termD_1\termD_2)_\delta =   \int_0^1 \dint{y} \int_0^u \dint{u'} \sum_{n_1,n_2} 
                \frac{q}{\sqrt{8\pi u'}}       
 \exp\left(-\frac{q^2}{2u'} (n_1-n_2)^2 -2q(y+(n_1+n_2))\right) \\
\times
        \left[
                e^{-qm} \indicator{(0,1)}[\sigma-y]\delta_{m,n_1+n_2} 
                + e^{-q(1-m)} \indicator{(-1,0)]}[\sigma-y]\delta_{m-1,n_1+n_2}
                \right]
\ .
\end{eqnarray*}         
using $\indicator{(0,1)}(y+1)=\indicator{(-1,0)}(y)$.
This term can in turn be split into three parts, each accounting for one of the
Kronecker $\delta$-functions and the possible values of $m$. To prevent
exponential suppression, $m=0$ is required for the prefactor $e^{-qm}$ 
and $m=0,1$ for the prefactor $e^{-q(1-m)}$, which gives rise to three
terms,
$(\termD_1\termD_2)_{\delta}=(\termD_1\termD_2)_{\delta_1}+(\termD_1\termD_2)_{\delta_2}+(\termD_1\termD_2)_{\delta_3}$.

In the first term $(\termD_1\termD_2)_{\delta_1}$ we have $m=0$ and 
$n_1=-n_2$ from the Kronecker $\delta$. 
Ensuring the indicator function is non-zero, we change the integration
limits of $y$ and using $\delta_{0,n_1+n_2}$ we write $n_1=-n_2=n$,
\begin{eqnarray*}
\fl
(\termD_1\termD_2)_{\delta_1}
=
        \int_0^\sigma \dint{y} \int_0^u \dint{u'} \sum_{n} 
                \frac{q}{\sqrt{8\pi u'}} \exp\left(-\frac{q^2}{u'} (2n^2) -2qy\right) \\
=
        \int_0^u \dint{u'} \sum_{n} \frac{1}{2q} \exp\left(-\frac{q^2}{u'} (2n^2)\right) \frac{q}{\sqrt{8\pi u'}} \left[ 1-e^{-2q\sigma} \right]
\ .
\end{eqnarray*}
In the limit of large $q$, 
the first exponential will asymptotically vanish unless $n=0$.
Since $\lfloor u'/q\rfloor=m=0$, we change the upper integration limit of $u'$ to
$\min(q,u)$, and apply $n=0$:
\begin{eqnarray*}
(\termD_1\termD_2)_{\delta_1} 
&=&
\int_0^{\min(1,\frac{u}{q})} q \ \dint{\sigma} \sum_{n} \frac{1}{2q} \exp\left(-\frac{q^2}{u'} (2n^2)\right) \frac{q}{\sqrt{8\pi q}} \frac{1}{\sqrt{\sigma}}\left[ 1-e^{-2q\sigma} \right] \\
&=&
        \frac{1}{4}\sqrt{\frac{q}{2\pi}} \int_0^{\min(1,\frac{u}{q})} \dint{\sigma} \frac{1}{\sqrt{\sigma}}\left[ 1-e^{-2q\sigma} \right]
\\
&=&
        \frac{1}{4}\sqrt{\frac{q}{2\pi}} \left[
                2\min\left(1,\frac{u}{q}\right) - \frac{1}{\sqrt{2q}}\Gamma \left( \frac{1}{2}, 2q\min\left(1,\frac{u}{q}\right) \right)
        \right]
\end{eqnarray*}
using $u'=\sigma q$ and
\begin{equation*}
\int_0^y \dint{x}\  x^\mu e^{-x} 
= \Gamma(\mu+1,y) 
\ ,
\end{equation*}
which converges to $\sqrt{\pi}$ for $\mu=1/2$ in the limit of large $y$.

Similar arguments can be used for $(\termD_1\termD_2)_{\delta_2}$. In
that case, the
support of the indicator function changes
from $[0,1]$ to $[-1,0]$ since $m=1$ and therefore the integral changes to:
\begin{eqnarray*}
(\termD_1\termD_2)_{\delta_2}
&=&
        \int_\sigma^\infty \dint{y} \int_0^u \dint{u'} \sum_{n} 
                \frac{q}{\sqrt{8\pi u'}} \exp\left(-\frac{q^2}{u'} (2n^2) -2qy\right) \\
&=&
        \frac{1}{4}\sqrt{\frac{q}{2\pi}} \left[
                \frac{1}{\sqrt{2q}}\Gamma \left( \frac{1}{2}, 2q\min\left(1,\frac{u}{q}\right) \right)
        \right]
\ .
\end{eqnarray*}

Finally, the contribution of $(\termD_1\termD_2)_{\delta_3}$, with $m=0$, is
\begin{eqnarray*}
\fl 
(\termD_1\termD_2)_{\delta_3}
=
        \int_0^1 \dint{y} \int_0^u \dint{u'} \sum_{n_1,n_2} 
                \frac{q}{\sqrt{8\pi u'}}       \\
\times  \exp\left(-\frac{q^2}{2u'} (n_1-n_2)^2 -2q(y+(n_1+n_2))\right)
                e^{-q} \indicator{(-1,0)}(\sigma-y)
\ .
\end{eqnarray*}
Since $m=0$, this means that $n_1+n_2=-1$. Letting $n_1=n$, this gives
$n_1-n_2=2n+1$ and thus
\begin{eqnarray*}
\fl
(\termD_1\termD_2)_{\delta_3}
=
        \int_0^1 \dint{y} \int_0^u \dint{u'} \sum_{n}
                \frac{q}{\sqrt{8\pi u'}}        
                \exp \left(
                    -q\left (\frac{q}{2u'} (2n+1)^2 +2y -1\right) 
                    \right) \\
    \times     e^{-q} \indicator{(-1,0)}(\sigma-y)  \ .
\ .
\end{eqnarray*}
Each exponential in the sum
vanishes for large $q$ unless
\begin{equation}
\frac{q}{2u'} (2n+1)^2 +2y -1  \leq 0 \ .
\end{equation}
Since $m=0$, $u'=q\sigma<q$, and from the indicator function follows $ y
>\sigma$, so that
\begin{equation}
0 \geq \frac{1}{2\sigma} (2n+1)^2 +2y -1 \geq \frac{1}{2\sigma} + 2y -1 
 > \frac{1}{2\sigma} + 2\sigma -1
\end{equation}
which implies $-2\sigma\ge (2\sigma-1)^2$ for $\sigma\ge0$, which is
impossible. The $(\termD_1\termD_2)_{\delta_3}$ term therefore vanishes.

In summary we have
\begin{eqnarray}
\fl\elabel{DD_summary}
(\termD_1\termD_2)_{\delta} = (\termD_1\termD_2)_{\delta_1} + (\termD_1\termD_2)_{\delta_2}+(\termD_1\termD_2)_{\delta_3}
=
        \frac{1}{2}\sqrt{\frac{q}{2\pi}} \left[
                \min\left(1,\frac{u}{q}\right) \right]
\ .
\end{eqnarray}
For $u$ small and by taking the limit $q\rightarrow\infty$, it is obvious that:
\begin{equation}
\lim_{q\rightarrow\infty} (\termD_1\termD_2)_{\delta} \rightarrow 0 
\elabel{DD_qfirst}
\end{equation}
Yet, taking $u\to\infty$ first gives $\min\left(1,\frac{u}{q}\right)=1$ and we
find to leading order in $q$:
\begin{equation}
\lim_{u\rightarrow\infty} (\termD_1\termD_2)_{\delta} 
= \frac{\Gamma^2}{v} q^{-1} \frac{1}{2}\sqrt{\frac{q}{2\pi}}
= \frac{\Gamma^2}{v}\frac{1}{2\sqrt{2\pi q}}
\elabel{DD_ufirst}
\end{equation}

Next we consider the roughness contribution of the 1-term, i.e. $y_1\neq y_2$: \\
\begin{eqnarray*}
\fl
(\termD_1\termD_2)_{1} 
=
        \int_0^1 \dint{y_1}
                \int_0^1 \dint{y_2} 
                \int_0^u \dint{u'}
                        \sum_{n_1,n_2} 
                                \frac{q}{\sqrt{8\pi u'}} 
                \exp
         \Bigg( -\frac{q^2}{2u'}
        \left[
                \frac{1}{2} (y_1-y_2) + (n_1-n_2)]
        \right]^2 \\
 -2q \left( \frac{1}{2} (y_1+y_2) + (n_1+n_2) \right) \Bigg) 
        \Bigg[
        e^{-qm} \indicator{(0,1)}\left[-\frac{1}{2} (y_1+y_2) + \sigma\right]\delta_{m,n_1+n_2} \\
 + e^{-q(m-1)} \indicator{(-1,0)}\left[-\frac{1}{2} (y_1+y_2) + \sigma\right]\delta_{m-1,n_1+n_2}
        \Bigg]
\end{eqnarray*}
This term, again, can be split into three parts,
$(\termD_1\termD_2)_1 = (\termD_1\termD_2)_{1_1} +
(\termD_1\termD_2)_{1_2} + (\termD_1\termD_2)_{1_3}$. Using similar as
for
the $\delta$-term, $(\termD_1\termD_2)_{1_1}$ contribution vanishes for large $q$ unless $m=0$, which 
implies $n_1=-n_2=n$. An SPA applied to the integral over $y_1$ gives a
minimum at
\begin{equation*}
y_{1_0} = y_2-\frac{4u'}{q}-4n
\ ,
\end{equation*}
so that the exponent is
\begin{equation*}
f(y_{1_0}) = q(2y_2-\frac{2u'}{q}-4n)
\ .
\end{equation*}
in the notation introduced above. The first term therefore gives
\begin{eqnarray*}
\fl
(\termD_1\termD_2)_{1_1} =
                \int_0^1 \dint{y_2} \sum_{n}
                \int_0^u \dint{u'}       \frac{q}{\sqrt{8\pi u'}} \sqrt{\frac{\pi}{q^2/8u'}} \ 
 \exp{ \left( -q\left(2y_2-\frac{2u'}{q}-4n\right) \right)}\\
\times                     
                \indicator{(0,1)}\left[ \frac{-1}{2} \left(y_2-\frac{4u'}{q}-4n+y_2 \right)+\sigma \right] 
                \indicator{(0,1)}\left[ y_2 -\frac{4u'}{q}-4n \right] \\
=
        \int_0^1 \dint{y_2} \int_0^u \dint{u'} \ 
                \exp{(-q(2y_2 - 2\sigma - 4n))} \\
 \times   \indicator{(0,1)}[ -y_2 + 3\sigma + 2n ]
                \indicator{(0,1)}[ y_2 - 4\sigma - 4n]
\end{eqnarray*}
Both indicator functions suggest $n=0$ or $n=-1$, otherwise the $(\termD_1\termD_2)_{1_1}$ term will not contribute. 
In the limit of large $q$, 
the term will be exponentially suppressed for $n=-1$ so that the only
case to be considered is $n=1$. However, using $n=0$ in both indicator
functions implies $\sigma<0$ for this term to contribute which
contradicts $\sigma\ge0$.
We conclude that $(\termD_1\termD_2)_{1_1}$ does not contribute. \\

By observation, $(\termD_1\termD_2)_{1_2}$ is identical to
$(\termD_1\termD_2)_{1_1}$ with the domain of the indicator function
changing from $[0,1]$ to $[-1,0]$ as $m=1$.
\begin{eqnarray*}
\fl
(\termD_1\termD_2)_{1_2} 
= 
                \int_0^1 \dint{y_2} \sum_{n}
                \int_0^u \dint{u'}       \frac{q}{\sqrt{8\pi u'}} \sqrt{\frac{\pi}{q^2/8u'}} \ \exp{ \left( -q\left(2y_2-\frac{2u'}{q}-4n\right) \right)}\\
 \times \                      
                \indicator{(-1,0)}\left[ \frac{-1}{2} \left(y_2-\frac{4u'}{q}-4n+y_2 \right)+\sigma \right] 
                \indicator{(0,1)}\left[ y_2 -\frac{4u'}{q}-4n \right] \\
=
        \int_0^1 \dint{y_2} \int_0^u \dint{u'} \ 
              \exp{( -q(2y_2- 2m - 2\sigma - 4n)) } \\
 \times \
                \indicator{(-1,0)}[ -y_2 + 3\sigma + 3m + 2n]
                \indicator{(0,1)}[ y_2 - 4\sigma - 4m - 4n] \\
=
        \int_0^1 \dint{y_2} \int_0^u \dint{u'} \ 
                \exp{ \left( -q\left(2y_2 - 2 - 2\sigma - 4n\right) \right)} \\
\times \ 
                \indicator{(-4,-3)}[ -y_2 + 3\sigma + 2n]
                \indicator{(4,5)}[ y_2 - 4\sigma - 4n]
\ .
\end{eqnarray*}
Both indicator functions contribute only if $n=-2$, in which case,
however, the 
term vanishes exponentially in large $q$.

Finally, we consider the $(\termD_1\termD_2)_{1_3}$ contribution, for
which $m=0$ and $n_1+n_2=-1$. By writing $n=n_1=-1-n_2$ and applying SPA
to the integral over $y_1$ we find the minimum of the exponent at
\begin{equation*}
y_{1_0} = y_2-\frac{4u'}{q}-2(2n+1) = y_2 - 4\sigma - 2(2n+1)
\end{equation*}
so that it becomes
\begin{equation*}
f(y_{1_0}) = q(2y_2 - \frac{2u'}{q} - 4n - 1) = q(2y_2 -2\sigma - 4n - 1)
\ .
\end{equation*}
and thus
\begin{eqnarray*}
\fl
(\termD_1\termD_2)_{1_3} 
= 
                \int_0^1 \dint{y_2} \sum_{n}
                \int_0^u \dint{u'}       \frac{q}{\sqrt{8\pi u'}} \sqrt{\frac{\pi}{q^2/8u'}} 
 \exp{ ( -q(2y_2 - 2\sigma -4n - 1) )}\\
 \times \                      
                \indicator{(-1,0)}\left[ \frac{-1}{2} \left(y_2 - 4\sigma - 2(2n+1) + y_2 \right)+\sigma \right] 
           \indicator{(0,1)}[y_2 - 4\sigma - 2(2n+1)] \\
=        \int_0^1 \dint{y_2} \int_0^u \dint{u'} \
                \exp{(-q(2y_2 - 2\sigma - 4n - 1))} \\
       \times          \indicator{(-2,-1)}[ -y_2 + 3\sigma + 2n]
                \indicator{(2,3)}[ y_2 - 4\sigma - 4n]
\ .
\end{eqnarray*}
Similar arguments as for $(\termD_1\termD_2)_{1_2}$ apply.
Both 
indicator functions contribute only if $n=-1$, 
where the term is exponentially suppressed. Therefore,
$(\termD_1\termD_2)_{1_3}$ does not contribute either.

Hence in total, considering both limits we have to leading order:
\begin{equation}
\lim_{q\rightarrow\infty} w^2_{\termD\termD} = 0\ \ \ \textrm{and}\ \ \ 
\lim_{u\rightarrow\infty} w^2_{\termD\termD} = \frac{\Gamma^2}{v}\frac{1}{2\sqrt{2\pi q}}
\elabel{DD_limit_summary}
\end{equation}
Any marginal cases are necessarily subleading compared to
\eref{DD_summary} and thus can be safely ignored.
 
By similar procedures, we obtain the leading order behaviour of 
$w^2_{\termC\termC}$ and $w^2_{\termC\termD}$ as follows:
\begin{equation}
\lim_{q\rightarrow\infty} w^2_{\termC\termC} = \frac{\Gamma^2}{v} \sqrt{\frac{u}{2\pi}}
\ \ \ \textrm{and}\ \ \ 
\lim_{u\rightarrow\infty} w^2_{\termC\termC} = \frac{\Gamma^2}{v} \frac{2}{3\sqrt{2\pi}} \sqrt{q}
\elabel{CC_limit_summary}
\end{equation}
and
\begin{equation*}
\lim_{q\rightarrow\infty} w^2_{\termC\termD} = 0
\ \ \ \textrm{and}\ \ \ 
\lim_{u\rightarrow\infty} w^2_{\termC\termD} = 0  \ .
\end{equation*}

In summary, the only non-zero contributions are \eref{DD_limit_summary} and
\eref{CC_limit_summary}.

\section*{References}
\bibliography{articles,books}

\end{document}